\documentclass[journal]{IEEEtran}
\usepackage{cite}
\usepackage{amsfonts}
\usepackage{mathrsfs}
\usepackage{graphicx}
\usepackage{amssymb}
\usepackage{amsmath}
\usepackage{amsthm}
\usepackage{bm}
\usepackage{url}
\usepackage{subfig}

\begin{document}

\title{Experimental Performance Evaluation of Cell-free Massive MIMO Systems Using COTS RRU with OTA Reciprocity Calibration and Phase Synchronization}
\DeclareRobustCommand*{\IEEEauthorrefmark}[1]{%
    \raisebox{0pt}[0pt][0pt]{\textsuperscript{\footnotesize\ensuremath{#1}}}}
\author{Yang Cao, Pan Wang, Kang Zheng, Xianghu Liang, Dongjie Liu, Mengting Lou, Jing Jin, Qixing Wang, Dongming Wang, Yongming Huang, Xiaohu You, Jiangzhou Wang
\thanks{Y. Cao, K. Zheng, X. Liang, D. Wang, Y. Huang and X. You are with National Mobile Communications Research Laboratory, Southeast University, Nanjing 210096, China (email: wangdm, xhyu@seu.edu.cn).
Y. Cao, P. Wang, D. Liu, D. Wang, Y. Huang and X. You are with Purple Mountain Lab., M. Lou, J. Jin, Q. Wang are with the China Mobile Research Institute, Beijing 100053,  China. J. Wang is with the School of Engineering, University of Kent, Canterbury CT2 7NT, U.K. (email: j.z.wang@kent.ac.uk).}
}

\maketitle

\begin{abstract}
Downlink coherent multiuser transmission is an essential technique for cell-free massive multiple-input multiple output (MIMO) systems,  and the availability of channel state information (CSI) at the transmitter is a basic requirement. To avoid CSI feedback in a time-division duplex system, the uplink channel parameters should be calibrated to obtain the downlink CSI due to the radio frequency circuit mismatch of the transceiver. In this paper, a design of a reference signal for over-the-air reciprocity calibration is proposed. The frequency domain generated reference signals can make full use of the flexible frame structure of the fifth generation (5G) new radio, which can be completely transparent to commercial off-the-shelf (COTS) remote radio units (RRUs) and commercial user equipments. To further obtain the calibration of multiple RRUs, an interleaved RRU grouping with a genetic algorithm is proposed, and an averaged Argos calibration algorithm is also presented. We develop a cell-free massive MIMO prototype system with COTS RRUs, demonstrate the statistical characteristics of the calibration error and the effectiveness of the calibration algorithm, and evaluate the impact of the calibration delay on the different cooperative transmission schemes.
\end{abstract}


\section{Introduction}
Spectral efficiency (SE) is one of key parameter indicators in the design of cellular mobile communication systems. For the fifth generation (5G) system with massive multiple-input multiple-output (MIMO)
technology \cite{Marzetta_mMIMO}, the SE can be increased to more than 50 bps/Hz. By using  the remote radio units (RRUs) deployed
in the existing cellular systems and introducing a coordinated multipoint (CoMP) transmission technique, the SE can be further increased. The related technologies are also referred to as distributed MIMO, cooperative MIMO, multiple transmission and reception points (multi-TRP), or cell-free massive MIMO (CF-mMIMO) \cite{you_6g_new}. CF-mMIMO can be viewed as an evolution of distributed MIMO
or CoMP \cite{Wang2020Implementation}. It employs scalable implementation to achieve coordinated multiuser transmission, thereby substantially improving SE \cite{You2021Distributed, bornson_scal, Ngo2017}. Therefore, CF-mMIMO has been considered as an enabling technology for the sixth generation (6G) \cite{you_6g_new}.

In the CF-mMIMO system, coherent downlink transmission is adopted to achieve space-division multiplexing, which usually relies on known downlink channel state information (CSI) at the central processing unit (CPU).
Exploiting the reciprocity of the over-the-air (OTA) channels in a time-division duplex (TDD) system, the feedback of the downlink CSI can be avoided, and then the signaling overhead can be reduced. However, in a practical system, the overall channels are composed of OTA propagation coefficients and the transmission coefficients introduced by the radio frequency (RF) transceivers. Since the RF transmitter and receiver have different transmission coefficients, the overall uplink and downlink channels are not reciprocal in practice. The mismatches in downlink-uplink channels include amplitude and phase mismatches. \cite{Palacios} showed that the downlink performance loss of a CF-mMIMO system is large when the phase mismatch is greater than 15$^\circ$. Actually, one of the factors  delaying the application of the CoMP technique in  commercial fourth generation (4G) and 5G systems was the reciprocity calibration or phase synchronization between distributed RRUs. A working group of 3GPP is studying the coherent transmission of multi-TRP for 5G new radio (NR) release 18 \cite{Samsung}, and the reciprocity calibration (or phase synchronization in some literature) of the downlink-uplink channels is a critical issue. Therefore, OTA reciprocity calibration is a crucial technique in future 6G-oriented CF-mMIMO \cite{Jiayi_review,Hussein}.

There are two main types of calibration methods, namely, hardware calibration and OTA calibration. The former requires an additional reference antenna, while the latter does not require extra hardware. Both methods have been extensively studied in TDD massive MIMO \cite{Shepard2012,JiangXiwen_TWC2018,Xiliang}. However, unlike in the centralized massive MIMO system, multiple RRUs are physically deployed at different locations in a CF-mMIMO system; therefore, OTA calibration is desirable. OTA calibration can be achieved by transmitting known reference signals (RSs) between RRUs or between the RRUs and the user equipments (UEs). The former is referred to as self-calibration, and the latter is named  UE-assisted calibration. Both algorithms can obtain the calibration coefficients of the RRUs, whereas self-calibration is  preferable since it is transparent {\footnote{In this paper, 'transparency' means that a commercial UE (or COTS RRU) can benefit from a coherent precoding with TDD calibration without being aware of how it is accomplished.}} to the UE. With uplink CSI and the calibration coefficients obtained from the collected calibration signals, coherent transmission can be achieved by using the calibrated downlink precoding \cite{Wence2015}.

OTA reciprocity calibration has been widely studied for distributed MIMO. \cite{Rogalin} investigated the calibration of multiple remote radio units (RRUs) and proposed a cluster-based calibration method. To date, distributed MIMO has been experimentally studied in WiFi/  long-term evolution (LTE) networks. \cite{Horia} and \cite{Katabi} implemented WiFi-based distributed MIMO, in which AirSync used the out-of-band signals for synchronization and MegaMIMO used a master device with multiple slave devices for synchronization. In \cite{Hamed}, a hierarchical synchronization architecture was proposed for the phase synchronization of the whole network. The design in  \cite{Hamed} is compatible with a 5G small-cell and has been verified on an LTE system. However, the method in \cite{Hamed} requires the deployment of virtual UEs to support calibration. \cite{Magounaki} and \cite{Magounaki_OAI5G} implemented a distributed MIMO system using the open-air-interface platform and proposed a master-slave calibration algorithm and a fast calibration method with RRU grouping, which can complete the calibration within 20 millisecond. In \cite{Yuan}, a distributed MIMO system was implemented, and a weighted least squares calibration method was proposed with CSI feedback from the UE.

For 5G evolution or 6G systems, the calibration signals should be standardized to achieve interoperability between devices. In addition, further research on  ultrafast, low-complexity and low-overhead calibration methods is needed to  attain scalable networking for CF-mMIMO. Furthermore, there are no publicly available performance evaluations of experimental systems for CF-mMIMO with scalable precoding and imperfect calibration.
The main research contributions of this study are summarized as follows:
\begin{enumerate}
  \item We propose a group-based fast OTA reciprocity calibration scheme with a genetic algorithm-aided RRU grouping and a 5G NR compatible calibration reference signal (CARS). Different from \cite{Hamed,Magounaki,Magounaki_OAI5G,Yuan}, the calibration process does not require UE feedback, transparently supports both commercial off-the-shelf (COTS) RRUs and commercial UEs and is especially suitable for the RRU implemented following the specification of open radio access network \cite{ORAN}. The CARS can achieve the calibration of up to 64 antennas on four orthogonal frequency-division multiplexing (OFDM \cite{Zhu_OFDM_1,Zhu_OFDM_2}) symbols in one slot (for example, approximately 134 $\mu$s for a 30 kHz subcarrier spacing).
  \item This study proposes an improved Argos calibration method suitable for CF-mMIMO, which can make full use of the wireless links between multiple RRUs to improve the calibration accuracy and reduce the calibration complexity.
  \item In this study, a CF-mMIMO prototype system with the 5G COTS RRU is developed, which is fully compatible with 5G commercial UEs. With the prototype UEs, the calibration coefficients obtained from the experimental system are analyzed and evaluated, including the statistical characteristics of the calibration errors and the performances of different calibration algorithms. Based on the testbed, we present the performance evaluations of the centralized and distributed precodings for downlink CF-mMIMO.
\end{enumerate}

The paper is organized as follows: Section II presents the 5G NR compatible CARS, the group-based calibration algorithm is investigated in Section III, and Section IV presents the experimental verification results, followed by the conclusion of this study in Section V.

The notation conventions in this paper are as follows: Matrices and vectors are denoted by bold italic uppercase and lowercase letters, respectively; ${\rm diag}({\bf x})$ is a diagonal matrix with $\bf x$ on its diagonal; ${\rm diag}({\bf A})$ denotes a vector with the main diagonal of $\bf A$; $(\cdot)^{\rm H}$, $(\cdot)^{\rm T}$, and $(\cdot)^{*}$ represent Hermitian transpose, transpose, and conjugate, respectively; $\odot$ represents Hadamard multiplication of two matrices; and $\oslash$ denotes elementwise division of two matrices.

\section{RRU Grouping and Design of a 5G NR-compatible calibration signal}

In a CF-mMIMO system, all of the RRUs should be calibrated for dynamic downlink coherent transmission. Location-based clustering is usually an effective way to reduce the calibration dimension for a CF-mMIMO system with a large number of RRUs. In \cite{Rogalin}, intercluster relative calibration and intracluster least squares calibration were proposed. An alternative transmission of the calibration signals was presented for the calibration of RRUs in a cluster. The method exhibits  an optimal performance but has a large calibration time.

In this study, the RRUs in a cluster are divided into two groups, and spatial-domain orthogonal calibration signals are transmitted between the two groups, thereby reducing the calibration time.
In this section, we study an optimal RRU grouping and then design a 5G NR-compatible calibration signal for the two groups.

\subsection{Interleaved RRU Grouping with Genetic Algorithm}
In \cite{Magounaki}, a group-based calibration was proposed. However, the authors did not  provide an effective method to obtain the optimal grouping.
Intuitively, to obtain a better calibration signal-to-noise ratio (SNR), the two groups of RRUs should be interleaved together as much as possible.
Note that when the RRUs are deployed, the calibration SNR between RRUs is mostly related to the large-scale fading, which is the relative distance between the RRUs.
To achieve a better performance of the group-based calibration, we minimize the sum of the distances between the two groups of RRUs. Therefore, we formulate
the minimization problem as follows:
\begin{align}\label{obj_fun}
&\min \sum\limits_{p \in {{\cal P}}} {\sum\limits_{q \in {{\cal Q}}} {{d_{p,q}}} } \\
&{\rm{s.t.}}\quad {{\cal P}} \cup {{\cal Q}} = {{\cal T}}~{\rm and }~{{\cal P}} \cap {{\cal Q}} = \emptyset~{\rm and }~\left| {\left| {{\cal P}} \right| - \left| {{\cal Q}} \right|} \right| \le 1 \nonumber
\end{align}
where ${{\cal T}}$ is a set of all RRUs, the antennas are divided into two sets ${{\cal P}}$ and ${{\cal Q}}$, in which there are $M$ antennas in  group ${{\cal P}}$ and $N$ antennas in  group ${{\cal Q}}$, and ${{d_{p,q}}}$ is the distance between  antenna $p$ in  group ${{\cal P}}$ and  antenna $q$ in  group ${{\cal Q}}$. Note that to simplify the problem, the difference in the number of RRUs in the two sets is less than or equal to one. With this constraint, the numbers of RRUs in the two groups are balanced.

When the number of RRUs is not large, we can use exhaustive search to obtain the optimal solution to the optimization problem (\ref{obj_fun}). However, when the number of RRUs is large, the complexity is high. Fortunately, the grouping is performed just once after deployment of the RRUs. This optimization problem can be described by binary variables and then solved by a genetic algorithm (GA). A genetic algorithm is a heuristic search inspired by the process of natural selection, which is commonly used to generate high-quality solutions to optimization and search problems by relying on bioinspired operators such as mutation, crossover and selection \cite{msrinivas_adaptive_1994,jzhang_clustering_2007}. Since the GA is a classical algorithm to solve the above problem, we will not give a detailed implementation here.

After the RRU grouping, we can send multiantenna orthogonal pilots to each other between the two groups, enabling a fast calibration. However, current 5G standards do not explicitly support RRUs sending signals to each other. Fortunately, we can take advantage of the dynamic configurable time slots of 5G NR to achieve this functionality.

\subsection{Slot configuration for calibration of RRU groups}

\begin{figure}
  \centering
  \includegraphics[width=3.3in]{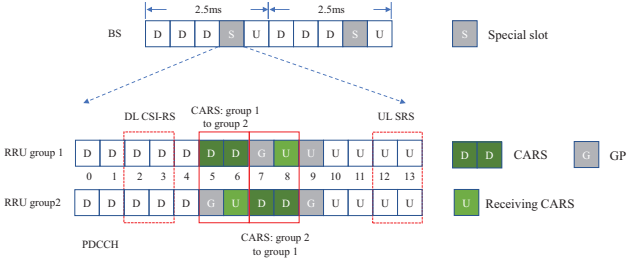}
  \caption{Configuration of a special slot for OTA reciprocity calibration. D, U and G denote the states in the OFDM symbol for the transmitting, receiving and guard periods, respectively.}
  \label{fig1}
\end{figure}

Figure \ref{fig1} shows a frame configuration with a period of $2.5 $ms. The special slot (S-slot) is well designed to perform a calibration of the two RRU groups. The S-slot configuration of Group 1 is a conventional pattern with a single guard period (GP) but the configuration of Group 2  has two GPs. Because of the different positions of the downlink-uplink switching points in the S-slots for the two RRU groups, we can transmit and receive reference signals between RRU groups. As shown in the figure, RRU Group 2 is in the GP and receiving state when the 5th and 6th symbols of the S-slot of RRU Group 1 are transmitting a CARS, and similarly, RRU Group 1 is in the receiving state when the 7th and 8th symbols of RRU Group 2 are transmitting a CARS. Therefore, we can achieve the transmission and reception of calibration signals between the two groups. Note that the CPU should not schedule UEs on these symbols. In addition, since 5G NR supports the configuration of the dynamic frame structure through downlink control information, we can configure S-slots as mentioned above according to the calibration period. For the normal slot, all the RRUs can be configured with a common S-slot to avoid cross-link interference.

To evaluate the calibration performance, we also insert the downlink CSI-RS and the uplink sounding reference signal (SRS) into the S-slot to measure the downlink-uplink CSI between the CPU and the UEs.

\subsection{CARS design considering uplink timing advance}

\begin{figure}[ht]
  \centering
  \includegraphics[width=1.8in]{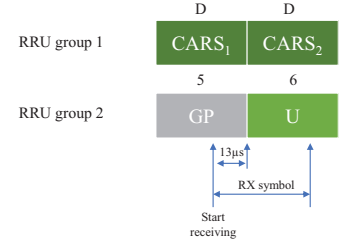}
  \caption{The effect of the uplink receiving advance  of the base station.}
  \label{fig2}
\end{figure}

In the 5G-NR system, considering the propagation delay and the TDD switching time, \cite{3GPP} requires an uplink timing advance; as a result, the uplink and downlink are staggered by approximately 13 $\mu$s (for the Sub6 GHz band). As shown in Figure \ref{fig2}, the RRUs in Group 2 start to receive 13 $\mu$s in the GP in advance. A commercial UE knows the timing advance, whereas the commercial RRUs in Group 1  have their own timing. If the RRUs in Group 1 just transmit CARS in the 6th symbol,  RRU Group 2  cannot receive the correct signal. Therefore, the two groups of RRUs are misaligned when transmitting signals to each other with the configuration of the S-slot in Figure \ref{fig1}.

Note that when the baseband processor has the capability to process time-domain OFDM symbols and adjust the TDD switch point, we can transmit and receive CARS in GP. However, for some commercial RRUs, such as those using the Option 7-2 standard \cite{NGMN}, the RRU has the functions of low-level physical layer processing, including fast Fourier transform (FFT )/inverse fast Fourier transform (IFFT), cyclic prefix (CP) addition and removal, and phase compensation. Therefore, this type of COTS RRU only receives the downlink frequency-domain signals from the baseband unit and transmits the uplink frequency-domain signals to the baseband unit. Therefore, it is necessary to design a frequency-domain calibration signal that considers the standardization issue.

\begin{figure}[ht]
  \centering
  \includegraphics[width=3.2in]{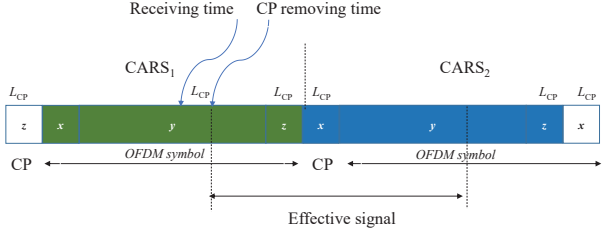}
  \caption{Two-symbol CARS.}
  \label{fig3}
\end{figure}

To ensure that the CPU receives a correct calibration signal, we propose a two-symbol frequency-domain CARS. A set of multiantenna orthogonal reference signals is designed according to the number of antennas in the two RRU groups. The frequency-domain reference signal CARS$_1$ for the $i$th antenna is expressed as
\begin{align}\label{cars_x1}
{\bf{x}}_i = {\left[ {{x_{i,1}},{x_{i,2}}, \ldots ,{x_{i,{N_{{\rm{FFT}}}}}}} \right]^{\rm{T}}},
\end{align}
where ${N_{{\rm{FFT}}}}$ is the length of the FFT, and the $n$-th element of the corresponding CARS$_2$ is given by
\begin{align}\label{cars_x2}
{\tilde x_{i,n}} = {x_{i,n}}\exp \left( {\iota \frac{{2\pi {L_{{\rm{CP}}}}}}{{{N_{{\rm{FFT}}}}}}n} \right),
\end{align}
That is, each sample undergoes an $L_{\rm CP}$  phase shift. After IFFT, and an addition of a CP, the two symbols have the characteristics shown in Figure \ref{fig3}; that is, the valid data of the former symbol is the CP of the ${N_{{\rm{FFT}}}}$  samples of the latter symbol.

With the above design, the RRUs start to receive the calibration signal in the first time-domain symbol with a receiving length of $N + L_{\rm CP}$; then, after removing the CP, a complete reference signal can be obtained. Note that there is a certain shift between the received calibration sequence and the original calibration sequence, and we can recover it by the phase rotation in the frequency domain. Therefore, the design can be completely transparent to the RRU in  Option 7-2 format \cite{NGMN}. With the CARS shown in Figure \ref{fig3}, we have a complete OFDM symbol at the receiver for calibration.

Commercial RRUs with multiple antennas for small-cell systems generally do not have internal calibration, so all the antennas of the RRUs should transmit calibration signals. When there is a large number of RRUs, the calibration signals should be orthogonal to obtain optimal channel estimations. Taking the SRS in 5G NR as an example, one OFDM symbol can support up to 16 antenna ports. Even 32 antenna ports can be multiplexed considering that the coverage area of small cells is usually small, and that the calibration coefficients vary little in the frequency domain. Therefore, with the slot configuration shown in Figure \ref{fig1}, we can achieve the calibration of a total of 64 antennas in two sets on four OFDM symbols. Considering  a subcarrier spacing of 30 kHz, the time consumption of the calibration is approximately 134 $\mu$s.

Finally, the CARS can be reused between clusters by using a similar idea of pilot reuse \cite{Liu,Wenbo}, and the calibration between the clusters can be implemented according to \cite{Rogalin}.

\section{Group-based calibration method}

To further improve the calibration accuracy with low complexity, in this section, we study the group-based calibration algorithms, including the traditional total least-square (TLS) and an improved Argos algorithm.

\subsection{TLS calibration algorithm}

Based on the calibration signals transmitted to each other, each receiver first performs a channel estimation to obtain the frequency-domain channel matrices between the two RRU groups. The channel matrices on a subcarrier between the two RRU groups are denoted as ${{\bf{H}}}_1$ and ${{\bf{H}}}_2$, which can be modeled as follows:
\begin{align}\label{H1}
{{\bf{H}}_1} = {{\bf{C}}_{{\rm{rx}},2}}{\bf{H}}{{\bf{C}}_{{\rm{tx}},1}},
\end{align}
\begin{align}\label{H2}
{{\bf{H}}_2} = {{\bf{C}}_{{\rm{rx}},1}}{{\bf{H}}^{\rm T}}{{\bf{C}}_{{\rm{tx}},2}},
\end{align}
where ${\bf{H}}$ is the OTA channel matrix between RRU Group 2 and RRU Group 1; ${{\bf{C}}_{{\rm{rx}},1}}$ and ${{\bf{C}}_{{\rm{tx}},1}}$ are the RF mismatch coefficients of the receiving and transmitting RRUs in Group 1, respectively; and ${{\bf{C}}_{{\rm{rx}},2}}$  and ${{\bf{C}}_{{\rm{tx}},2}}$ are the RF mismatch coefficients of the receiving and transmitting RRUs in Group 2, respectively, each of which is modeled as a diagonal matrix. The following calibration matrices are defined:
\begin{align}\label{C1}
{{\bf{C}}_{1}} = {\bf{C}}{_{{\rm{rx}},1}}{\bf{C}}_{{\rm{tx}},1}^{ - 1},
\end{align}
\begin{align}\label{C2}
{{\bf{C}}_{2}} = {\bf{C}}{_{{\rm{rx}},2}}{\bf{C}}_{{\rm{tx}},2}^{ - 1}.
\end{align}
These are the calibration coefficients of the RRUs in Group 1 and Group 2. According to the above calibration matrix, the following equation is true:
\begin{align}\label{eq_h_c}
{{\bf{H}}_1}{{\bf{C}}_{1}} = {{\bf{C}}_{2}}{\bf{H}}_2^{\rm{T}}.
\end{align}
The calibration vectors are defined as
$${{\bf{c}}_{1}} = {\rm{diag}}\left( {{{\bf{C}}_{1}}} \right), \quad {{\bf{c}}_{2}} = {\rm{diag}}\left( {{{\bf{C}}_{2}}} \right),$$
\[{{\bf{c}}_1} = {\left[ {\begin{array}{*{20}{c}}
{{c_{1,1}}}& \cdots &{{c_{1,M}}}
\end{array}} \right]^{\rm{T}}},\]
\[{{\bf{c}}_2} = {\left[ {\begin{array}{*{20}{c}}
{{c_{2,1}}}& \cdots &{{c_{2,N}}}
\end{array}} \right]^{\rm{T}}},\]
Then, the calibration vector of all RRUs can be expressed as
$${{\bf{c}}_{{\rm{cal}}}} = {\left[ {{\bf{c}}_{1}^{\rm{T}},{\bf{c}}_{2}^{\rm{T}}} \right]^{\rm{T}}}.$$

According to (\ref{eq_h_c}), in the presence of noise, we can establish the following TLS optimization objective function \cite{Kaltenberger,Rogalin}:
\begin{align}\label{eq_optim_obj}
&{\mathop {\arg {\rm{ min}}}\limits_{{{\bf{c}}_{{\rm{cal}}}}} } {\left\| {{{\bf{H}}_1}{{\bf{C}}_{1}} - {{\bf{C}}_{2}}{\bf{H}}_2^{\rm{T}}} \right\|^2}\nonumber\\
&{{\rm{s}}{\rm{.t}}{\rm{.}}}{\left\| {{{\bf{c}}_{{\rm{cal}}}}} \right\|^2} = 1
\end{align}
The calibration model described in (\ref{eq_optim_obj}) is the same as the UE-assisted calibration model proposed in \cite{Kaltenberger}. The above objective function can be expressed as
\begin{align}
& J\left( {{{\bf{c}}_{1}},{{\bf{c}}_{2}}} \right) = {\left\| {{{\bf{H}}_1}{{\bf{C}}_{1}} - {{\bf{C}}_{2}}{\bf{H}}_2^{\rm{T}}} \right\|^2}\nonumber\\
&{\rm{ = Tr}}\biggl( {\bf{C}}_{1}^{\rm{H}}{\bf{H}}_1^{\rm{H}}{{\bf{H}}_1}{{\bf{C}}_{1}} + {{\bf{C}}_{2}}{\bf{H}}_2^{\rm{T}}{\bf{H}}_2^{\rm{*}}{\bf{C}}_{2}^{\rm{H}} \nonumber\\
 &\quad\quad- {{\bf{C}}_{2}}{\bf{H}}_2^{\rm{T}}{\bf{C}}_{1}^{\rm{H}}{\bf{H}}_1^{\rm{H}} - {{\bf{H}}_1}{{\bf{C}}_{1}}{\bf{H}}_2^{\rm{*}}{\bf{C}}_{2}^{\rm{H}} \biggr).
\end{align}
It is noted that the following is true:
\[{\rm{Tr}}\left( {{\bf{C}}_{1}^{\rm{H}}{\bf{H}}_1^{\rm{H}}{{\bf{H}}_1}{{\bf{C}}_{1}}} \right) = {\bf{c}}_{1}^{\rm{H}}{\rm{diag}}\left[ {{\rm{diag}}\left( {{\bf{H}}_1^{\rm{H}}{{\bf{H}}_1}} \right)} \right]{{\bf{c}}_{1}},\]
\[{\rm{Tr}}\left( {{{\bf{C}}_{2}}{\bf{H}}_2^{\rm{T}}{\bf{H}}_2^{\rm{*}}{\bf{C}}_{2}^{\rm{H}}} \right) = {\bf{c}}_{2}^{\rm{H}}{\rm{diag}}\left[ {{\rm{diag}}\left( {{\bf{H}}_2^{\rm{H}}{{\bf{H}}_2}} \right)} \right]{{\bf{c}}_{2}},\]
\[{\rm{Tr}}\left( {{{\bf{C}}_{2}}{\bf{H}}_2^{\rm{T}}{\bf{C}}_{1}^{\rm{H}}{\bf{H}}_1^{\rm{H}}} \right) = {\bf{c}}_{1}^{\rm{H}}\left( {{\bf{H}}_1^{\rm{H}} \odot {{\bf{H}}_2}} \right){{\bf{c}}_{2}},\]
\[{\rm{Tr}}\left( {{{\bf{H}}_1}{{\bf{C}}_{1}}{\bf{H}}_2^{\rm{*}}{\bf{C}}_{2}^{\rm{H}}} \right) = {\bf{c}}_{2}^{\rm{H}}\left( {{\bf{H}}_2^{\rm{H}} \odot {{\bf{H}}_1}} \right){{\bf{c}}_{1}}.\]
Therefore,
\[J\left( {{{\bf{c}}_{1}},{{\bf{c}}_{2}}} \right) = {\bf{c}}_{{\rm{cal}}}^{\rm{H}}{\bf{A}}{{\bf{c}}_{{\rm{cal}}}},\]
where
\[{\bf{A}} = \left[ {\begin{array}{*{20}{c}}
{{{\bf{A}}_{11}}}&{{{\bf{A}}_{12}}}\\
{{{\bf{A}}_{21}}}&{{{\bf{A}}_{22}}}
\end{array}} \right],\]
\[{{\bf{A}}_{11}} = {\rm{diag}}\left[ {\rm{diag}}\left( {\bf{H}}_1^{\rm{H}}{{{\bf{H}}_1}} \right) \right],\]
\[{{\bf{A}}_{22}} = {\rm{diag}}\left[ {\rm{diag}}\left( {\bf{H}}_2^{\rm{H}}{{{\bf{H}}_2}}  \right)\right],\]
\[{{\bf{A}}_{12}} =  - {{\bf{H}}_2} \odot {\bf{H}}_1^{\rm{H}},\]
\[{{\bf{A}}_{21}} = {\bf{A}}_{12}^H =  - {{\bf{H}}_1} \odot {\bf{H}}_2^{\rm{H}}.\]

Then, the optimization objective function can be rewritten as
\begin{align}\label{eq_optim_obj2}
&{\mathop {\arg {\rm{ min}}}\limits_{{{\bf{c}}_{{\rm{cal}}}}} } ~~{\bf{c}}_{{\rm{cal}}}^{\rm{H}}{\bf{A}}{{\bf{c}}_{{\rm{cal}}}}\nonumber\\
&{{\rm{s}}{\rm{.t}}{\rm{.}}}{\left\| {{{\bf{c}}_{{\rm{cal}}}}} \right\|^2} = 1
\end{align}
According to \cite{Rogalin}, the optimal solution of ${{\bf c}_{{\rm{cal}}}}$ is the eigenvector corresponding to the minimum eigenvalue of ${\bf{A}}$, so that we may obtain the calibration coefficients of all the RRUs.

Note that the TLS calibration requires an eigenvalue decomposition of $\bf A$ with the computation complexity $O\left[ {{{\left( {M + N} \right)}^3}} \right]$.

\subsection{Averaged Argos calibration algorithm}\label{sec_3b}

To reduce the complexity of implementation, we propose an improved Argos calibration. For simplicity of the following description, the channel noise is neglected.
Based on the two sets of channel matrices, we have the following:
\[{{\bf{\Theta }}_1} = {{\bf{H}}}_{1} \oslash {{\bf{H}}}_2^{\rm{T}},\]
where $\oslash$ represents the division of the corresponding elements of the two matrices. Therefore, the above equation can be expressed as:
\[{{\bf{\Theta }}_1} = {{{\bf{c}}}_{2}}{\left[ {{\rm{diag}}\left( {{{\bf{C}}}_{1}^{ - 1}} \right)} \right]^{\rm{T}}} = \left[ {\begin{array}{*{20}{c}}
{\frac{{{c_{2,1}}}}{{{c_{1,1}}}}}&{\frac{{{c_{2,1}}}}{{{c_{1,2}}}}}& \cdots &{\frac{{{c_{2,1}}}}{{{c_{1,M}}}}}\\
{\frac{{{c_{2,2}}}}{{{c_{1,1}}}}}&{\frac{{{c_{2,2}}}}{{{c_{1,2}}}}}& \cdots &{\frac{{{c_{2,2}}}}{{{c_{1,M}}}}}\\
 \vdots & \vdots & \vdots & \vdots \\
{\frac{{{c_{2,N}}}}{{{c_{1,1}}}}}&{\frac{{{c_{2,N}}}}{{{c_{1,2}}}}}& \cdots &{\frac{{{c_{2,N}}}}{{{c_{1,M}}}}}
\end{array}} \right]\]
where ${\rm{diag}}\left( {{{\bf{C}}}_{1}^{ - 1}} \right)$ is the reciprocal of each element of the vector ${{\bf{c}}_{1}}$. The matrix ${{\bf{\Theta }}_1}$ is a rank-1 matrix. We define the following:
\begin{align}
{{{\bf{\Theta }}}_2} &= {{{\bf{H}}}_2} \oslash {{\bf{H}}}_1^{\rm{T}} = {{{\bf{c}}}_{1}}{\left[ {{\rm{diag}}\left( {{{\bf{C}}}_{2}^{ - 1}} \right)} \right]^{\rm{T}}} \nonumber\\
&= \left[ {\begin{array}{*{20}{c}}
{\frac{{{c_{1,1}}}}{{{c_{2,1}}}}}&{\frac{{{c_{1,1}}}}{{{c_{2,2}}}}}& \cdots &{\frac{{{c_{1,1}}}}{{{c_{2,N}}}}}\\
{\frac{{{c_{1,2}}}}{{{c_{2,1}}}}}&{\frac{{{c_{1,2}}}}{{{c_{2,2}}}}}& \cdots &{\frac{{{c_{1,2}}}}{{{c_{2,N}}}}}\\
 \vdots & \vdots & \vdots & \vdots \\
{\frac{{{c_{1,M}}}}{{{c_{2,1}}}}}&{\frac{{{c_{1,M}}}}{{{c_{2,2}}}}}& \cdots &{\frac{{{c_{1,M}}}}{{{c_{2,N}}}}}
\end{array}} \right]
\end{align}
where ${\rm{diag}}\left( {{\bf{C}}_{1}^{ - 1}} \right)$ is a diagonal matrix formed by the reciprocal of each element of the vector ${{\bf{c}}_{1}}$.

Consider the $N$th antenna of Group 2 as a reference antenna to describe the proposed calibration algorithm. The last column of ${{\bf{\Theta }}_2}$ is defined as ${\bm{\vartheta }}$.
Then, each column of ${{\bf{\Theta }}_2}$  is multiplied by the corresponding diagonal element of ${{\bf{\Theta }}_1}$ to obtain
$${{\bf{\hat \Theta }}_2} = {{\bf{\Theta }}_2}{\rm{diag}}\left( {{{\bf{\Theta }}_1}} \right).$$
Then, we construct the following:
\[{\bf{\Theta }} = \left[ {\begin{array}{*{20}{c}}
{{{{\bf{\hat \Theta }}}_2}}\\
{{{\bf{\Theta }}_1}}
\end{array}} \right]{\rm{diag}}\left( {\bm{\vartheta }} \right).\]

It can be seen that all columns of ${\bf{\Theta }}$ are equal to ${{{\bf c}_{{\rm{cal}}}}}/{{{c_{2,N}}}}$ except the last column containing all ones (in practice, the last row can be directly assigned to 1 and is not involved in the calculation). We can average the matrix by column to obtain the final calibration coefficient.

However, in practice, the channel estimation is not perfect, and especially in distributed MIMO systems, when the distance between two RRUs is too large, the calibration signal experiences severe fading, resulting in a poor SNR, which eventually affects the accuracy of the calibration. If some solutions deviate too much from the normal value, the averaging in the above algorithm inevitably leads to performance degradation, so some anomalous solutions must be removed. Therefore, the following operations are performed for ${{\bf{\Theta }}_1}$ and ${{\bf{\Theta }}_2}$:
\begin{enumerate}
  \item Find antenna $i$ in group $1$ that has the smallest sum of the distances from all antennas in Group 2, and find antenna $j$ in Group 2 that has the smallest sum of the distances from all antennas in group $1$.
  \item Find all the RRU pairs in RRU Group 1 and RRU Group 2 with an SNR less than a specified threshold. For example, assume that the SNR between antenna pair $\left( {p,q} \right)$ is less than a given threshold, where $p$ is an antenna in Group 1 and $q$ is an antenna in Group 2; then, let ${\left[ {{{\bf{\Theta }} _1}} \right]_{q,p}} = 0$  and ${\left[ {{{\bf{\Theta }} _2}} \right]_{p,q}} = 0$. Note that the zero-setting operation is not performed on the $j$th row of ${{\bf{\Theta }}_1}$ or the $i$th row of ${{\bf{\Theta }}_2}$.
\end{enumerate}

Based on the above results, matrices ${{\bf{\Theta }}_2}$  and ${{\bf{\Theta }}_1}$ are calibrated for the $j$th and $i$th rows, respectively,  and then averaged. The Argos algorithm for multiantenna averaging requires $2MN$ divisions, $2MN$ multiplications, and $MN$ additions for calibration, showing a lower complexity and   an easier hardware implementation than TLS.

\begin{figure}
  \centering
  \includegraphics[width=3in]{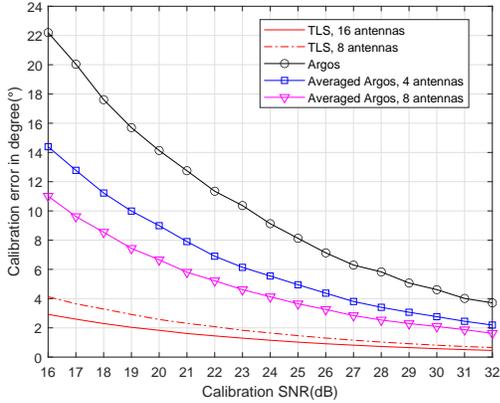}
  \caption{Performance comparisons of different calibration algorithms.}
  \label{tls_avags_cmp}
\end{figure}

\subsection{Simulation results}

Figure \ref{tls_avags_cmp} shows the phase errors (in degrees) of the calibration coefficients for different system configurations with different calibration algorithms. In the simulation, the amplitudes of the RF gains are assumed to be of log-normal distribution with  a 1 dB variance, and the phases are assumed to be of uniform distribution with range $(-\pi, \pi)$. In the simulation, there are two RRUs, each with 8 (or 16) antennas. The inter-RRU channel is modeled by an independent and identically distributed (i.i.d.) Rayleigh fading. Considering that the amplitude error of the calibration coefficient has a minor impact on overall performance \cite{Wei_scis,Palacios}, we are only concerned about phase error. To evaluate the accuracy of phase calibration, we normalize the calibration coefficients with respect to one antenna.

From Figure \ref{tls_avags_cmp}, it can be seen that TLS significantly outperforms Argos and has an average phase error less than 1$^\circ$ at a calibration SNR of 25 dB. Argos with a single reference antenna has poor calibration performance, with an error of approximately 4$^\circ$ at a calibration SNR of 31 dB. After averaging over multiple antennas, the calibration performance of Argos is improved; for example, an SNR gain of more than 8 dB is obtained for averaging eight antennas. Considering the complexity and accuracy, we can use the method of averaging over multiple antennas to improve the calibration accuracy of Argos in a practical system.

\begin{figure}
  \centering
  \includegraphics[width=3in]{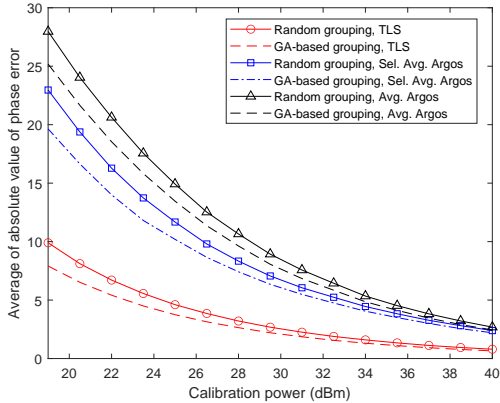}
  \caption{Performance comparisons for GA-based grouping and different calibration algorithms.}
  \label{fig_ga_grouping}
\end{figure}

Next, we will demonstrate the performance of GA-based grouping and the averaged Argos algorithm considering the practical deployment of RRUs. For large-scale fading, we utilize the following model \cite{Sanguinetti}:
\[\lambda \left( d \right) = 2\bar \lambda {\left[ {1 + {{\left( {{{1 + d} \mathord{\left/
    {\vphantom {{1 + d} {{d_0}}}} \right.
    \kern-\nulldelimiterspace} {{d_0}}}} \right)}^\alpha }} \right]^{ - 1}},\]
where $\bar \lambda $ denotes the path loss at the reference point, which is given by
\[\bar \lambda_{\rm dB}  =  - 34.5 - 20{\log _{10}}\left( {{d_0}} \right) - {N_{\rm NF}} - 10{\log _{10}}\left( {N_{\rm BW}} \right) - {N_0}.\]
The reference distance ${d_0}$ is 10 meters. ${N_{\rm NF}}$ denotes the noise figure, which is set to 9 dB, ${N_0}$ is the thermal noise power density with -174 dBm/Hz, and the system bandwidth ${N_{\rm BW}}$ is 1 MHz. The total number of RRUs is 8. Each RRU is with single antenna. The path loss exponent $\alpha$ is set to $3.7$. The RRU positions are randomly generated within a circle with a radius of 200 meters for 300 times; the corresponding small-scale fading is randomly generated 300 times for each.

From the simulation results in Figure \ref{fig_ga_grouping}, it can be seen that the performance of the TLS algorithm has been improved after a GA-based grouping. For a calibration power  of 26 dBm, the performance was improved by about 20$\%$ compared with the random grouping. It is also demonstrated that for the practical deployment direct averaging is not the best option. We can select appropriate channels with the method in subsection \ref{sec_3b} to obtain significant performance improvement.

\begin{figure}
  \centering
  \includegraphics[width=3in]{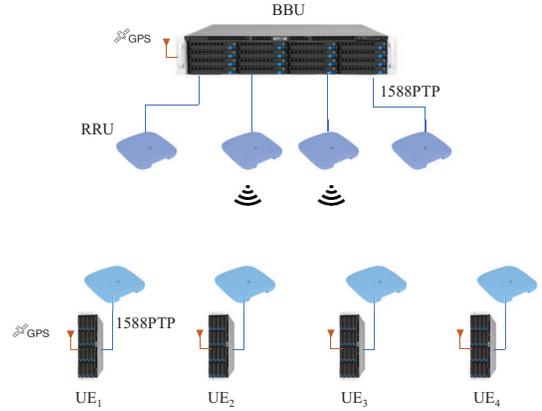}
  \caption{System configurations.}
  \label{fig_prototype_sys}
\end{figure}

\section{OTA calibration test results of a CF-mMIMO system}

\begin{figure*}[t]
  \centering
  \includegraphics[width=7in]{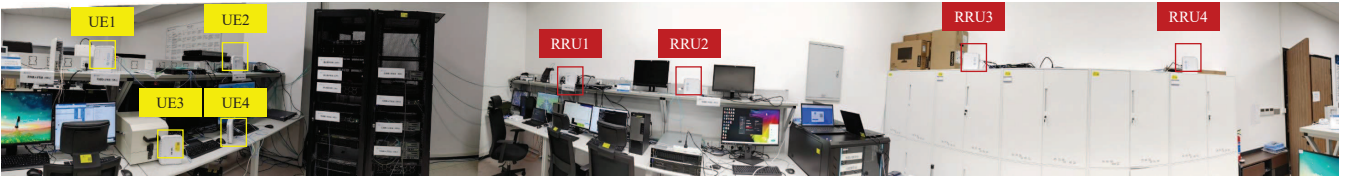}
  \caption{Prototype system.}
  \label{fig_prototype_rru}
\end{figure*}

\subsection{Prototype system}
Figure \ref{fig_prototype_sys} shows the CF-mMIMO prototype system. The test environment is shown in Figure \ref{fig_prototype_rru}. Low-cost RRUs for 5G indoor coverage are used. The system operates in the 4.9 GHz band with a bandwidth of 100 MHz. There are four RRUs and four UEs in the system, and the number of antennas of each RRU/UE is four. An evolved common public radio interface (eCPRI) is used between the RRU and the fronthaul accelerator card, which complies with the open radio access network. The CPU provides time synchronization and the reference clock for multiple RRUs by using IEEE 1588PTPv2 and SyncE protocols. Each RRU has an independent local oscillator (LO), but the four RF chains in one RRU share the same LO. The prototype UE uses the similar hardware platform as the CPU.

We divide the four RRUs on the CPU side into two groups and use the frame structure shown in Figure \ref{fig1}. The calibration signal between the two RRU groups is the same as the uplink SRS, which occupies 272 resource blocks (RBs). The channel estimation in this study adopts a Wiener interpolation based on the uniform power delay profile. For downlink channel estimation, we adopt the CSI-RS with 16 orthogonal ports.

In the practical system, it is difficult to obtain the perfect calibration coefficients of each RRU; consequently, a reasonable baseline is very important to evaluate the experimental performance. In the prototype system, for the UE-assisted calibration, we consider two RRU groups and 16 antennas per group, whereas for the self-calibration of the two RRU groups, there are 8 antennas per group. Obviously, as shown in Figure \ref{tls_avags_cmp}, UE-assisted calibration has better performance. Moreover, for the UE-assisted calibration, the CPU has both uplink and downlink channels, and then it is regarded as perfect CSI feedback. Therefore, we finally select UE-assisted TLS as the baseline.

In the following, we evaluate the experimental data in detail in terms of the time-frequency characteristics of the calibration coefficients, statistical characteristics of the calibration error, the performance of the calibration algorithms and CF-mMIMO downlink performance. Unless stated otherwise, the calibration algorithm is a TLS-based self-calibration.

\subsection{Time-frequency characteristics of the calibration coefficients}

%
%

\begin{figure}
\centering
\subfloat[]{\includegraphics[width=0.83\columnwidth]{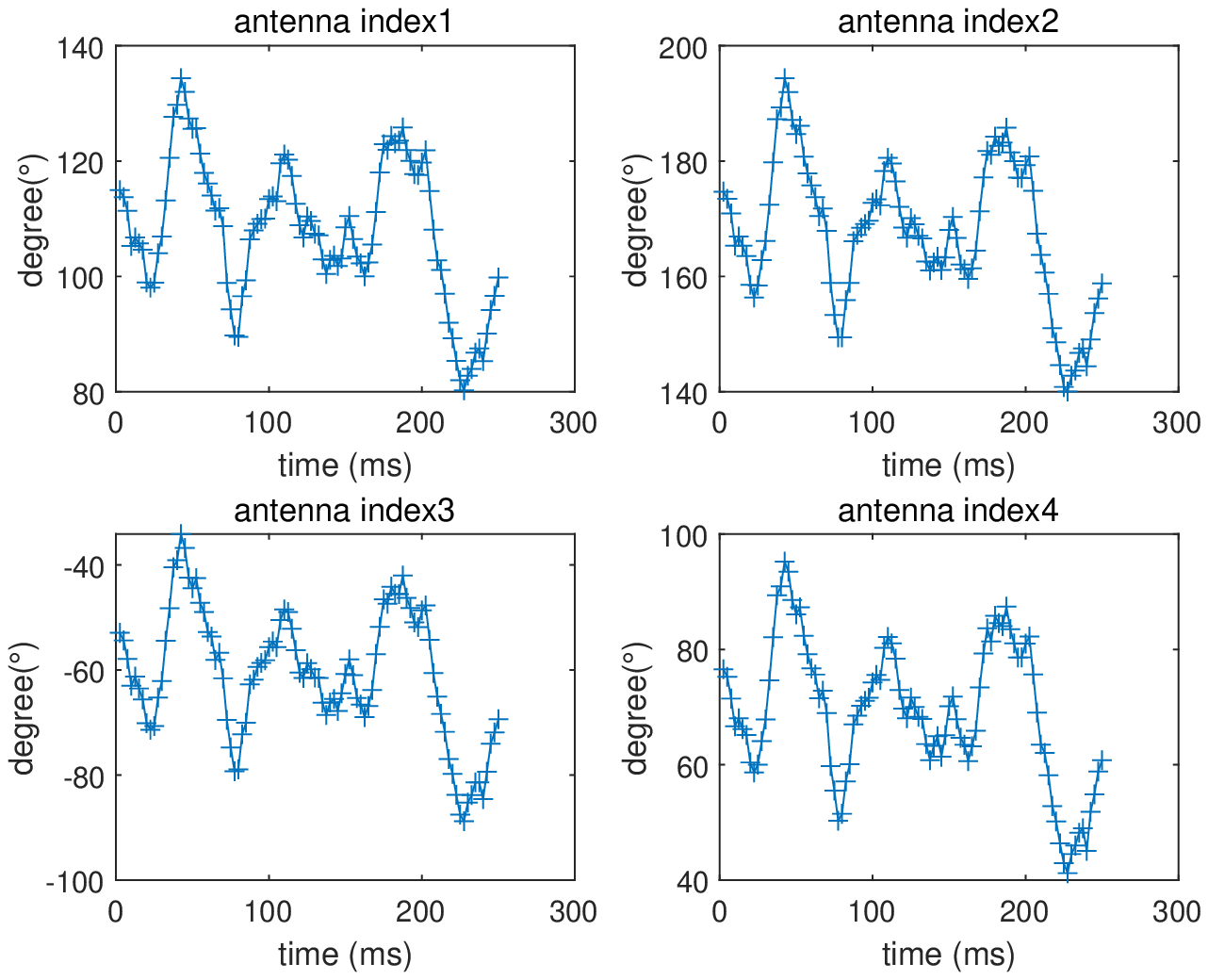}\label{fig_coef_tf_a}}
\\
\subfloat[]{\includegraphics[width=0.73\columnwidth]{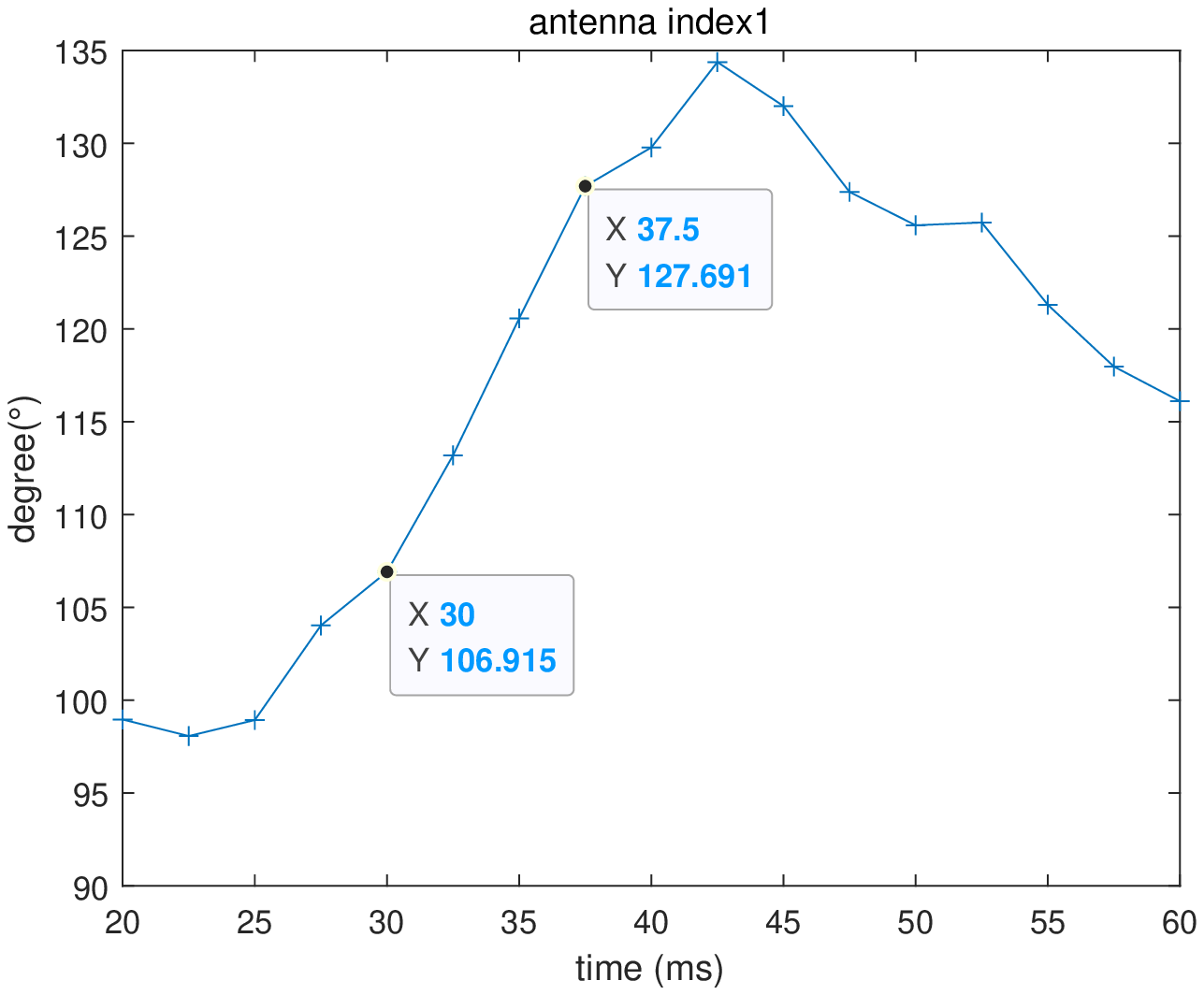}\label{fig_coef_tf_b}}
\\
\subfloat[]{\includegraphics[width=0.83\columnwidth]{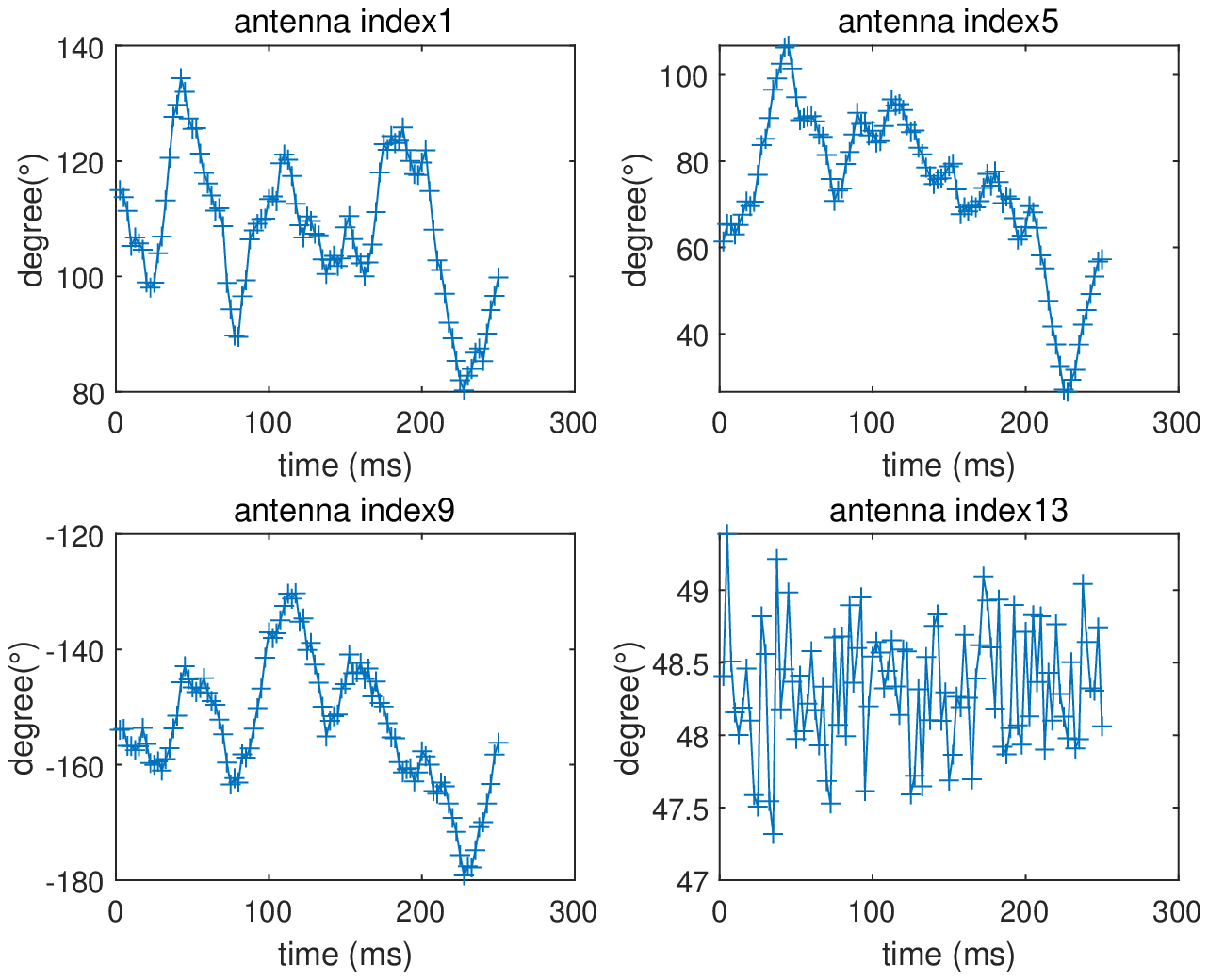}\label{fig_coef_tf_c}}
\phantomcaption
\end{figure}

\begin{figure}
\ContinuedFloat
\centering
\subfloat[]{\includegraphics[width=0.73\columnwidth]{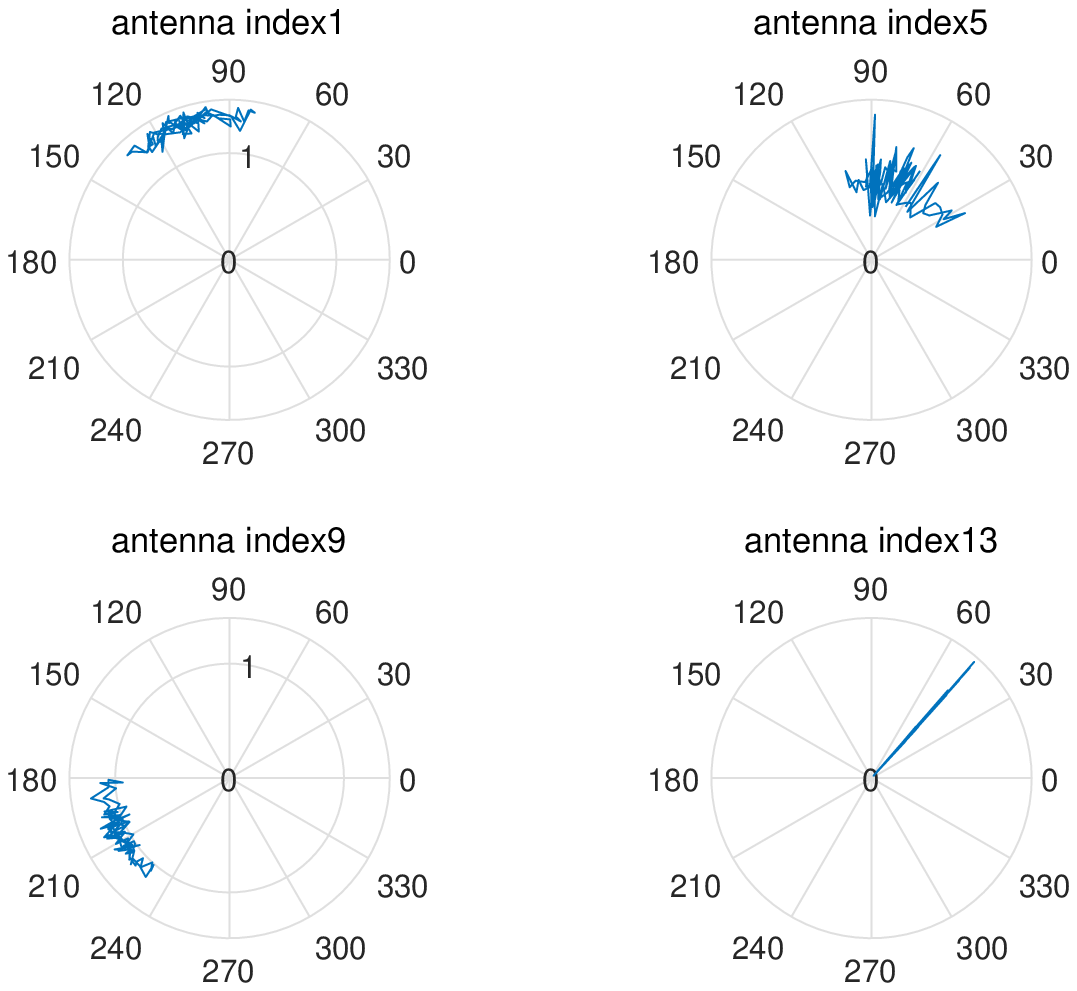}\label{fig_coef_tf_d}}
\\
\subfloat[]{\includegraphics[width=1\columnwidth]{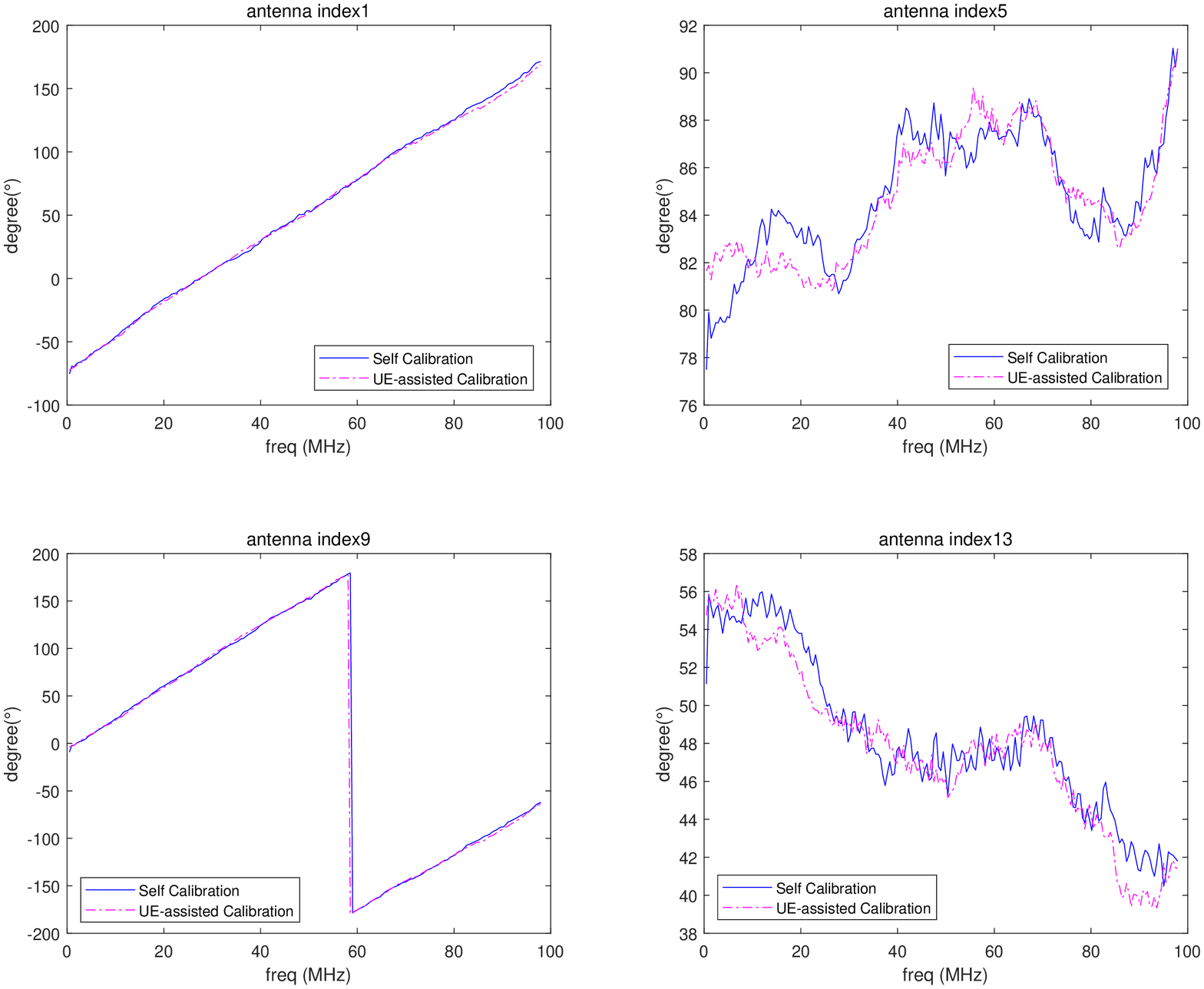}\label{fig_coef_tf_e}}
\\
\caption{Frequency-time domain characteristics of the calibration coefficients: (a) Time domain characteristics of the calibration coefficients of RRU 1 for a given subcarrier. (b) Time domain characteristics of the calibration coefficients of antenna 1 in RRU 1 for a given subcarrier. (c) Time domain characteristics of the calibration coefficients of antennas 1, 5, 9, and 13 in RRUs 1, 2, 3, and 4, respectively. (d) Time domain characteristics of the calibration coefficients under the polar candidate system of antennas 1, 5, 9, and 13 in RRUs 1, 2, 3, and 4, respectively. (e) Frequency domain characteristics of the calibration coefficients of antennas 1, 5, 9, and 13 in RRUs 1, 2, 3, and 4, respectively.}
\end{figure}

We divide the four RRUs in the system into two groups. The antennas of the two RRUs in Group 1 are numbered one to four for RRU$_1$ and five to eight for RRU$_2$, and the antennas of the two RRUs in Group 2 are numbered nine to twelve for RRU$_3$ and thirteen to sixteen for RRU$_4$. We normalize the calibration coefficients with respect to that of the last antenna, so that the 16th antenna has a calibration coefficient of 1.

Figure \ref{fig_coef_tf_a} shows the phase change of the calibration coefficient with time for the four antennas of RRU$_1$ on a certain subcarrier. Due to the LO phase drift, the phases of the calibration coefficients of the RRU on a given subcarrier drift with time, and the range is approximately -30$^\circ$ and +30$^\circ$ with respect to the center phase. Since the four channels in the RRU have a common LO, the phases of the calibration coefficients of the RRU have a fixed phase difference, and they are basically synchronous. The phase difference in the RRU remains unchanged over a long period of time (with an observation time of 250 ms). Normally, the phase difference is related to the environment and temperature, and varies by minutes. However, the calibration coefficient of an RF chain changes rapidly with time due to the LO phase drift. Figure \ref{fig_coef_tf_b} shows that at an interval of 7.5 ms, the LO drift exceeds 20$^\circ$, which will significantly degrade the performance of the coherent joint transmission, as demonstrated later.

Figure \ref{fig_coef_tf_c} shows the variation of the calibration coefficients with time for one antenna of each RRU. There is not only a fixed phase difference but also phase asynchrony between RRUs. In addition, since we use the 16th antenna as a reference, the phase of the calibration coefficient of the 13th antenna varies little, and the phase change is within $\pm 1.5^\circ$. Figure \ref{fig_coef_tf_d} shows the polar plot of the calibration coefficients for one antenna of each RRU. It is also seen that the phase rotation is approximately between -30$^\circ$ and +30$^\circ$ with respect to the center phase.

Next, we study the frequency domain performance of the calibration coefficients. The calibration signals are transmitted over the air, the delay between the transmission and reception is reciprocal, and theoretically, the calibration algorithm can eliminate the OTA timing delay. However, in the prototype system, although the RRUs recover the timing from 1588PTP packets, they still have very small timing differences. As shown in Figure \ref{fig_coef_tf_e}, the phase of the calibration coefficients basically varies linearly with the subcarrier. The results show that the calibration coefficients of both the UE-assisted calibration and the self-calibration involve a timing delay which is demonstrated as a linear phase shift in the frequency domain. In this paper, the timing delay is referred to the 16th antenna, and usually remains constant for a long time. The delays involved in the calibration coefficients of antenna five and thirteen are about 0.15 ns and -0.33 ns respectively, and the delays for antenna one and nine are close to one sample (about 8 ns). Fluctuations in the calibration coefficients of adjacent subbands are not easily observed for antennas 1 and 9. It can be seen from antennas 5 and 13 that there are also certain fluctuations in different subbands, for example, up to approximately 2$^\circ$ for six adjacent RBs (with a bandwidth of approximately 2 MHz).

\textbf{Remark 1}: With a common reference clock, there is no carrier frequency offset (CFO) for the RRUs in the prototype system{\footnote{Note that if the RRUs have different reference clocks, there will be a CFO between the different RRUs, and the calibration coefficients will encompass the CFO.}}. We can see that the phase of the calibration coefficient for a given subcarrier varies in the range of $\pm 30^\circ$ with a long-term constant value as the center. Due to the excellent coherence between the multiple channels of the RRU, the phase differences of the calibration coefficients are almost constant (the phase error is within $\pm 1.5^\circ$) for a long time. When the RRU is capable of self-calibration, for each RRU, we just need to estimate one calibration coefficient. Then, the overhead of the OTA calibration can be reduced, or with the same overhead, we can calibrate more RRUs.

\textbf{Remark 2}: The phase drift due to the phase-locked loop of each RRU is inevitable.  The phase drift is relatively large in one frame (10 ms). We may need to transmit CARS with a short period to obtain an accurate phase synchronization. The problem of LO phase drift is also related to the design of the LO phase-locked loop. With LO phase tracking \cite{McNeill,Koch}, we can reduce the overhead of the OTA calibration.

\textbf{Remark 3}: Considering that the calibration coefficients vary less over a few RBs, we can increase the subcarrier bandwidth of the CARS. For example, using a subcarrier spacing of 120 kHz, it is possible to put one CARS into a GP.

Limited by the current COTS RRU, the overhead of the CARSs in this study is still high. However, it is feasible to further reduce the overhead by considering the above discussion.

\subsection{Calibration error analysis}

Using the UE-assisted TLS calibration as a baseline, we compared the performance of the self-calibration in the CPU. Let the UE-assisted calibration coefficient be $X$ and the calibration coefficient of the self-calibration be $Y$; then, we define the following variables:
$$\theta  = {\rm{angle}}\left( {{Y/X}} \right), \quad \rho  = \left| {{Y/X}} \right|,$$
where $\theta$  and $\rho$ denote the phase and magnitude of the error, respectively.

\begin{figure}
\centering
\subfloat[]{\includegraphics[width=0.83\columnwidth]{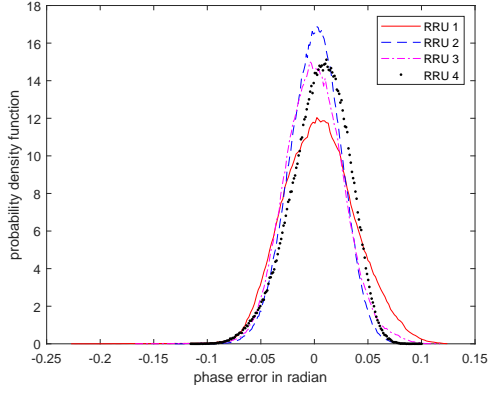}\label{fig_coef_pdf_a}}
\quad
\subfloat[]{\includegraphics[width=0.83\columnwidth]{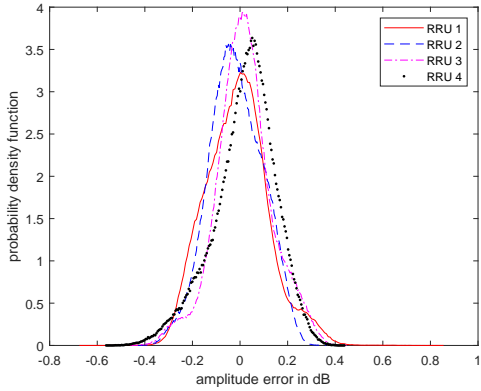}\label{fig_coef_pdf_b}}
\caption{Statistical characteristics of the calibration coefficients: (a) Phase PDF of the calibration error. (b) Amplitude PDF of the calibration error.}
\end{figure}

Considering that the calibration errors of the four antennas in each RRU have the same statistical characteristics, we treat them as one random variable. Figure \ref{fig_coef_pdf_a} shows the statistical characteristics of the calibration errors of the four RRUs. It can be seen that the phase of the calibration error approximately follows a normal distribution and that the amplitude can be approximated to a lognormal distribution; this is consistent with the theoretical result of \cite{weihao2016}. As   seen in Figure \ref{fig_coef_pdf_b}, the magnitude of the calibration error is small, less than 0.3 dB. Since the amplitude error is small and has little impact on the system performance, we focus on the statistical characteristics of the phase error in the following.

\begin{figure}
  \centering
  \includegraphics[width=3in]{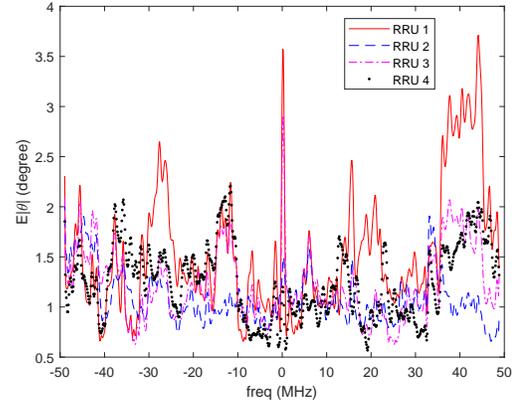}
  \caption{The phase calibration errors of the four RRUs on different subbands.}
  \label{fig_phase_error_rru}
\end{figure}

\begin{figure}
\centering
\subfloat[]{\includegraphics[width=0.83\columnwidth]{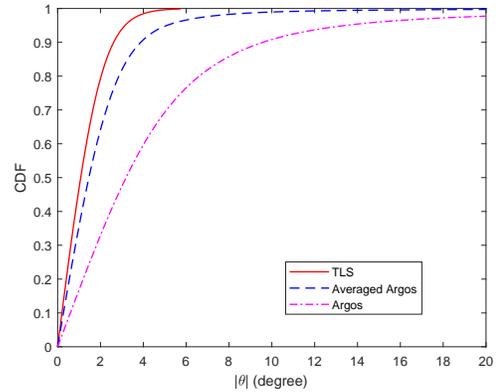}\label{fig_coef_delay_cdf_a}}
\quad
\subfloat[]{\includegraphics[width=0.83\columnwidth]{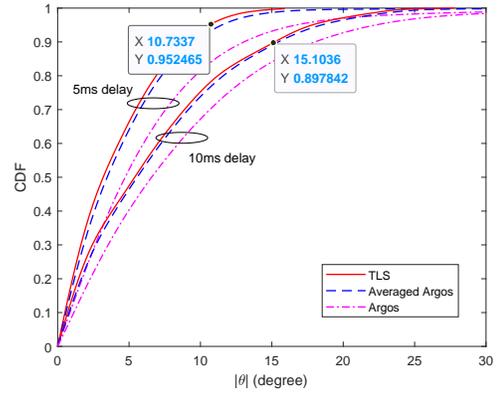}\label{fig_coef_delay_cdf_b}}
\caption{CDF of the absolute phase calibration errors: (a) calibration without delay. (b) Calibration with delay.}
\end{figure}

Figure \ref{fig_phase_error_rru} shows the performance of the calibration coefficient errors of the four RRUs on different subbands. We average the absolute values of the phases of the calibration errors on the same subcarrier and find that the maximum calibration error is approximately 4$^\circ$. The averaged absolute phase errors over frequency and time for the four RRUs are 1.53$^\circ$, 1.08$^\circ$, 1.23$^\circ$, and 1.25$^\circ$. From Figure \ref{tls_avags_cmp}, we can see that for an SNR of 19 dB, the performance gap between a TLS with eight antennas and 16 antennas is approximately 1$^\circ$. Then, if the calibration SNRs for UE -assisted calibration and self-calibration are the same, the averaged phase error for TLS-based self-calibration of RRU$_2$ is approximately 3$^\circ$. Figure \ref{fig_coef_delay_cdf_a} gives the cumulative distribution function (CDF) of the absolute values of the phases of the calibration errors for all antennas of all the RRUs. It can be seen that the proposed averaged Argos algorithm has much better performance than that of traditional Argos, and the performance is very close to the TLS. Then, the averaged Argos algorithm can achieve a good tradeoff between complexity and performance.

In practical systems, because of the signal processing delay or the large calibration period, the calibration coefficients calculated in the S-slot are usually applied to the downlink slots with a certain delay. Therefore, in Figure \ref{fig_coef_delay_cdf_b} we evaluate the CDF of the absolute value of the phase of the calibration error in the presence of the calibration delay. Even a delay of 5 ms causes a large phase error, with 5$\%$ of the channels having a phase error of more than 10$^\circ$, while a delay of 10 ms results in 10$\%$ of the channels having a calibration coefficient phase error of more than 15$^\circ$. In addition, the performance of the averaged Argos algorithm can approach that of TLS due to the calibration delay.

\textbf{Remark 4}: The calibration coefficients of the RRU change significantly with time due to phase drift of the independent LO of each RRU. Therefore, it is necessary to further evaluate the impact of calibration delay on system performance.

\subsection{Comparison of the SE performance of the system with 16 downlink data streams}
We first evaluate the total SE of transmitting 16 downlink data streams using joint zero-forcing (ZF) precoding. The CPU uses the calibration coefficients and the uplink channels to obtain the downlink channels and further calculate the ZF precoding. The signal to interference plus noise (SINR) of each data stream is calculated from the downlink equivalent channel to obtain the total SE. We consider the following four cases: UE-assisted TLS calibration (which can be regarded as the ideal downlink CSI feedback, referred to as UE-assisted TLS), TLS self-calibration (referred to as TLS), averaged Argos self-calibration (referred to as averaged Argos), and Argos self-calibration (referred to as Argos).

\begin{figure}
\centering
\subfloat[]{\includegraphics[width=0.83\columnwidth]{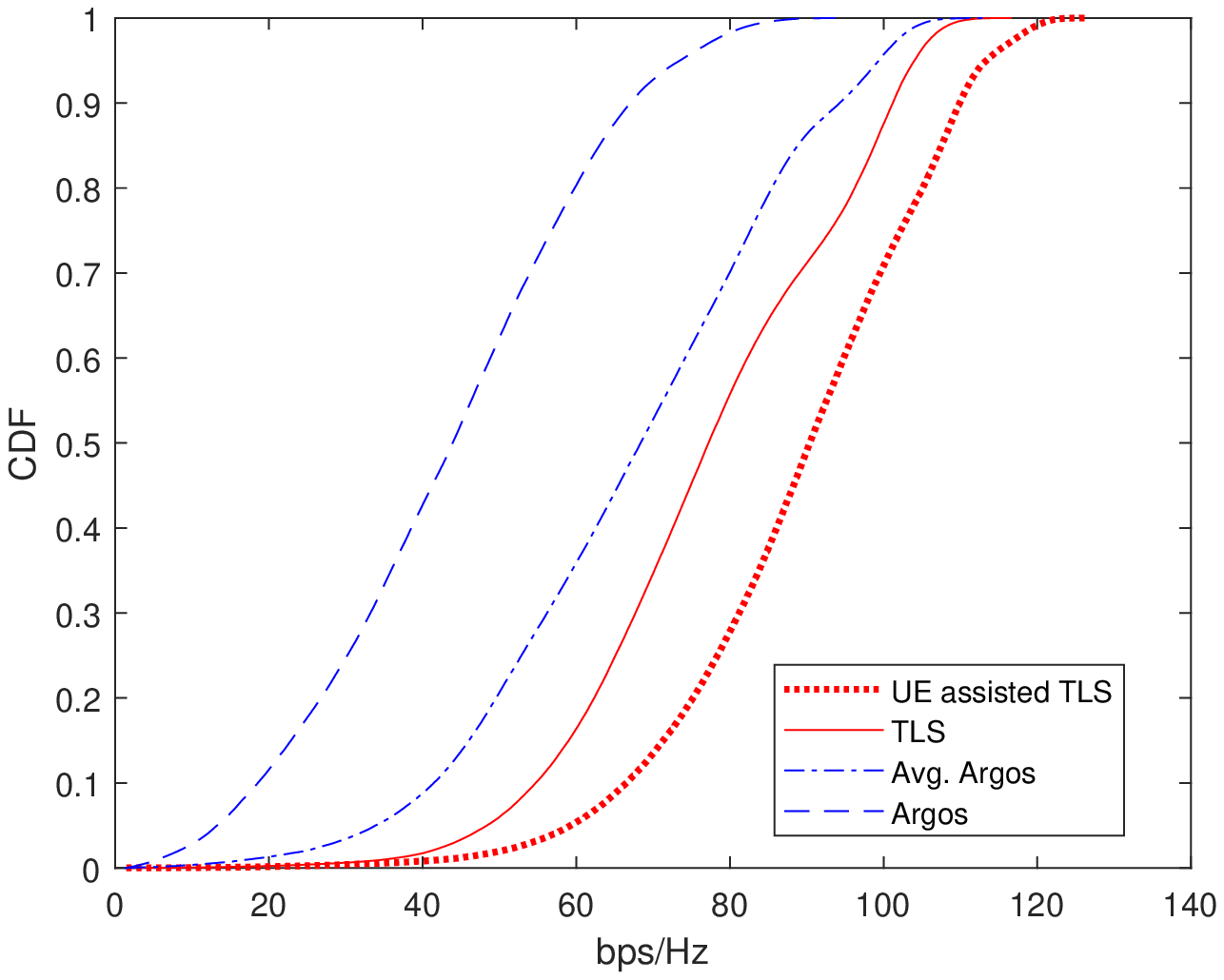}\label{fig_full_jp_cdf_a}}
\quad
\subfloat[]{\includegraphics[width=0.83\columnwidth]{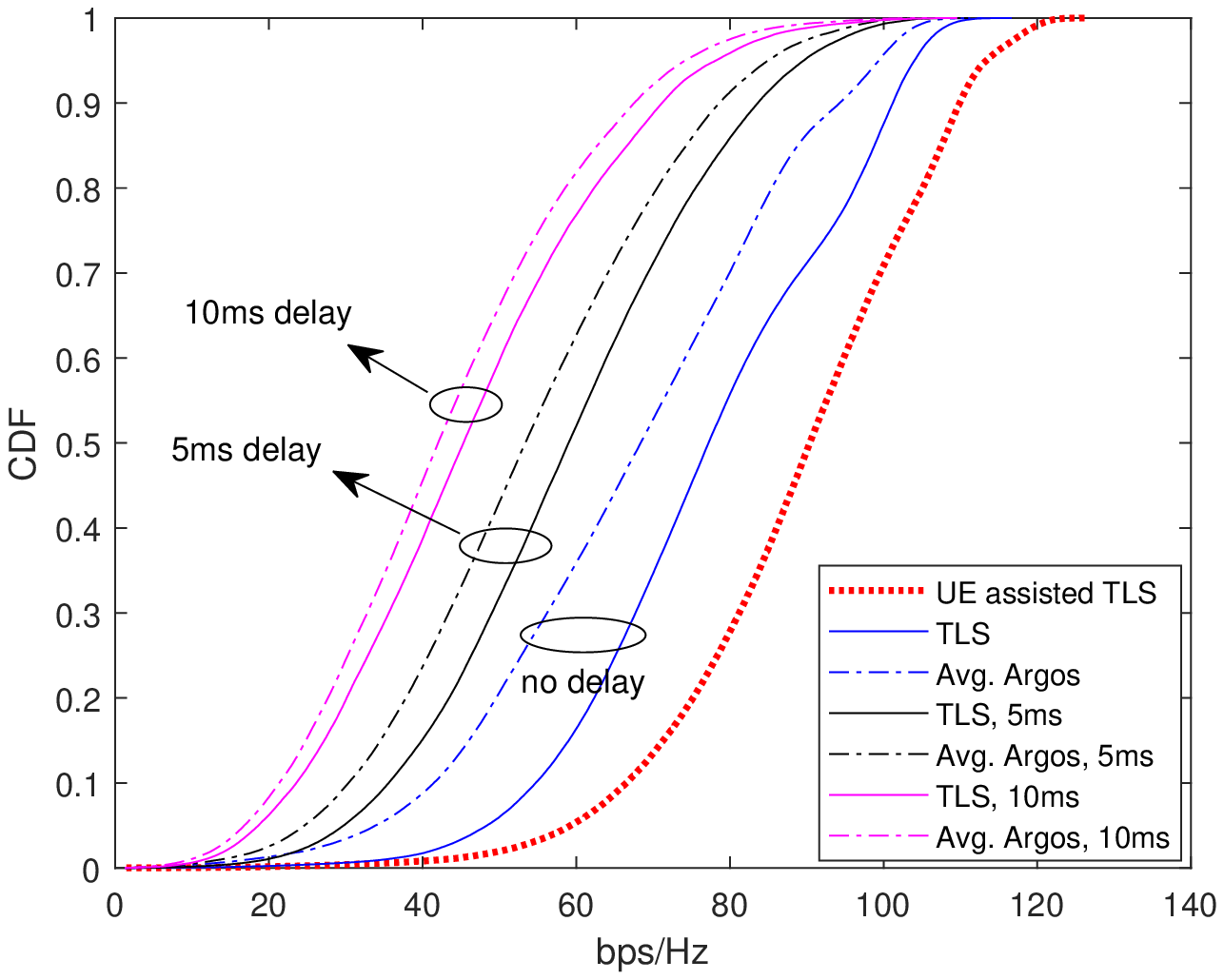}\label{fig_full_jp_cdf_b}}
\caption{SE of full cooperation: (a) calibration without delay. (b) Calibration with delay.}
\end{figure}

Figure \ref{fig_full_jp_cdf_a} shows the total SE of the 16 data streams without considering the calibration delay. With a 10$\%$ outage probability, compared to the UE-assisted TLS, the TLS, the averaged Argos, and the Argos algorithms have performance losses of 18$\%$, 37$\%$, and 72$\%$, respectively.

\begin{figure}
  \centering
  \includegraphics[width=3in]{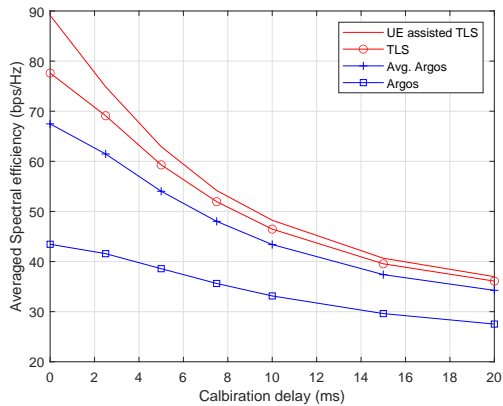}
  \caption{Average SE for different calibration delay.}
  \label{fig_full_jp_avg_se}
\end{figure}

Figure \ref{fig_full_jp_cdf_b} presents the total SE considering a calibration delay of 5 ms and 10 ms, under which even the performance loss of the TLS reaches 46$\%$ and 64$\%$, respectively, and the performance loss of the averaged Argos further increases. Figure \ref{fig_full_jp_avg_se} shows the relationship between the average SE and the calibration delay. As the calibration delay increases, the performance loss is significant due to the calibration error.

\textbf{Remark 5}: Cooperative downlink coherent multiuser transmission is susceptible to  calibration errors in the case of full spatial multiplexing, and even a small calibration delay has a significant impact on the system performance.

\subsection{Comparison of the SE performance with scalable precoding algorithms}
For CF-mMIMO, we usually assume that the number of UEs is much smaller than the number of antennas. Therefore, we evaluate the performance of the system using distributed precoding when one antenna is used for each prototype UE. Due to the poor performance of the Argos algorithm, in this subsection, we just show the SE performances of CF-mMIMO with the following three calibration algorithms: the UE-assisted TLS, the TLS and the averaged Argos. We consider three commonly used precoding schemes, namely, maximum-ratio transmission (MRT), local regularized zero forcing (L-RZF ) and joint processing-regularized zero forcing (JP-RZF).

\begin{figure}
  \centering
  \includegraphics[width=3in]{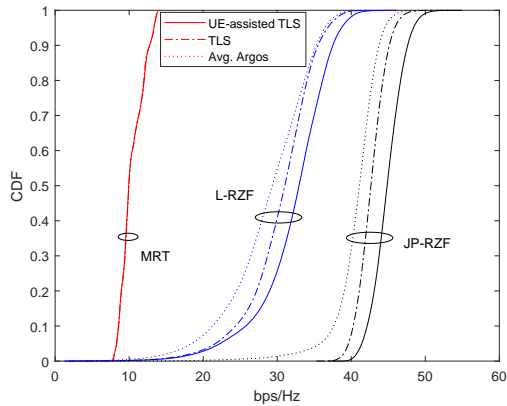}
  \caption{CDF of SE for different precoding schemes.}
  \label{fig_cf_lp_cdf_a}
\end{figure}

\begin{figure}
  \centering
  \includegraphics[width=3in]{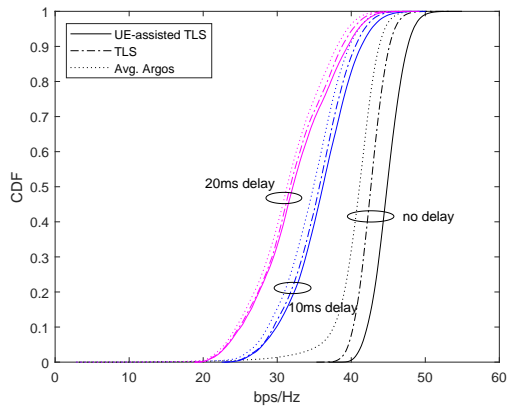}
  \caption{CDF of SE for JP-RZF with and without delays.}
  \label{fig_cf_lp_cdf_b}
\end{figure}

Figure \ref{fig_cf_lp_cdf_a} shows the CDF of the SE of the different precoding schemes without delay. Relatively speaking, the system performance is less affected by the calibration error due to the small number of spatial data streams. The performance of these calibration algorithms is nearly the same as that for MRT precoding. The calibration error has little impact on the performance of L-RZF and JP-RZF.
Figure \ref{fig_cf_lp_cdf_b} shows the impact of a calibration delay on JP-RZF. The 10-ms and 20-ms delays have a significant impact on the performance of JP-RZF. For the TLS, with an outage probability of 10$\%$, the 10-ms and 20-ms delays lead to performance losses of 29$\%$ and 41$\%$, respectively.

\begin{figure}
  \centering
  \includegraphics[width=3in]{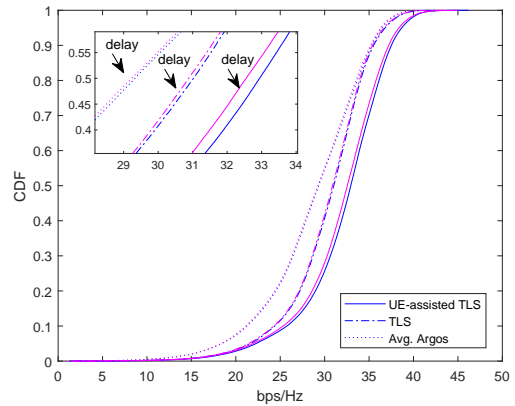}
  \caption{CDF of SE for L-RZF with 20 ms delays.}
  \label{fig_cf_lp_cdf_c}
\end{figure}

\begin{figure}
  \centering
  \includegraphics[width=3in]{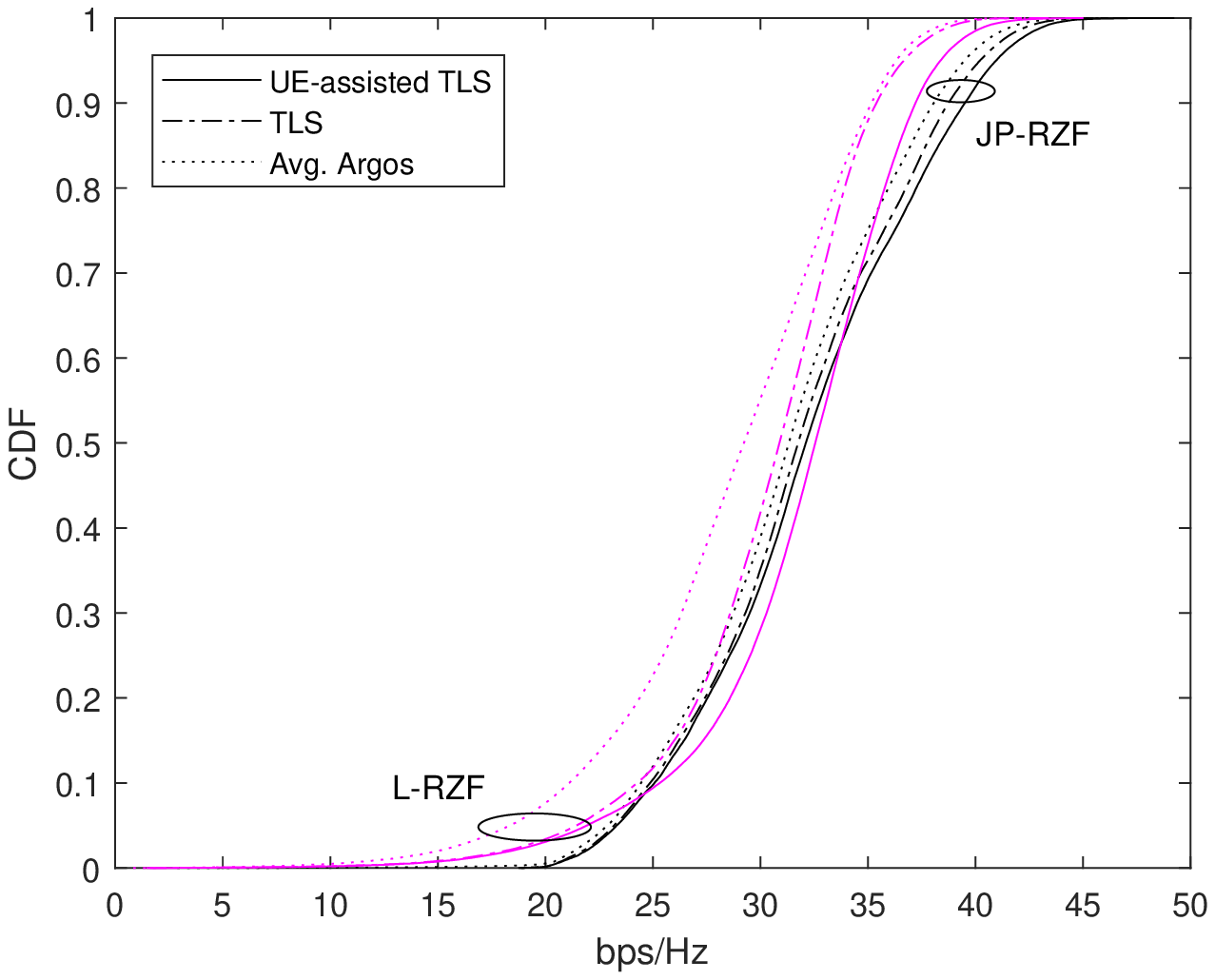}
  \caption{CDF of SE for JP-RZF and L-RZF with 20 ms delays.}
  \label{fig_cf_lp_cdf_d}
\end{figure}

Figure \ref{fig_cf_lp_cdf_c} shows the performance of L-RZF with a 20 ms calibration delay. Since each RRU is equipped with four antennas in the system, the common LO is realized, and the phase changes synchronously in the RRU, so that the L-RZF with four streams can successfully suppress the interstream interference. As has been pointed out (see Remark 1), the average value of the phase between RRUs remains constant for a long time, and the phase drift is in the range of (-$\pi$/6, $\pi$/6). In Appendix A, we prove that for  a small range of the phase drift, the performance loss of local precoding can be omitted. Therefore, the L-RZF is not sensitive to the phase drift of the calibration coefficients between RRUs. Figure \ref{fig_cf_lp_cdf_d} illustrates the impact of  a 20-ms calibration delay on L-RZF and JP-RZF, and there is little difference in the performances of the two in this case.

\begin{figure}
  \centering
  \includegraphics[width=3in]{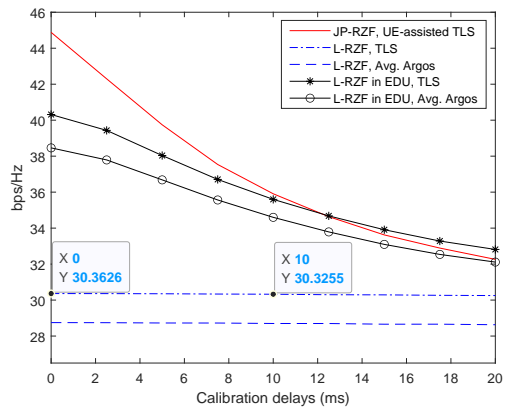}
  \caption{Average SE for different calibration delay.}
  \label{fig_cf_lp_avg_se}
\end{figure}

Figure \ref{fig_cf_lp_avg_se} illustrates the relationship between the average SE and the calibration delay. The performance of L-RZF is the worst but is nearly unaffected by the calibration delay. We also show the performance of a cell-free system with edge distributed units (EDUs) \cite{Xiaohu_IET}. We divide the four RRUs into two groups, with two RRUs in each group, and the RRUs are connected to the EDUs in which the uplink coordinated reception and the downlink L-RZF precoding are implemented. The use of this scheme leads to a better performance than that of L-RZF implemented in RRUs, and the EDU-based L-RZF even outperforms JP-RZF when the calibration delay is large. However, the L-RZF implemented by EDUs remains susceptible to  calibration errors.

\textbf{Remark 6}: With the L-RZF, when the number of downlink data streams is less than or equal to the number of antennas of each RRU, the system performance is robust to the calibration delay.

\section{Conclusions}
In this study, TDD OTA calibration and phase synchronization techniques for 6G-oriented CF-mMIMO have been investigated. First, an OTA reciprocity CARS compatible with the 5G frame structure was designed that is transparent to commercial UEs and RRUs and enables fast self-calibration of the RRUs. The averaged Argos calibration was proposed for a group-based calibration, which can achieve a good tradeoff between complexity and performance. We have developed a CF-mMIMO prototype platform based on 5G commercial COTS RRUs. Based on the testbed, the time-frequency characteristics of the calibration coefficients, the statistical characteristics of the calibration error, the performance of joint coherent processing and local precoding were studied. The experimental results showed that in the absence of a common LO for commercial RRUs, joint processing with multiple RRUs is extremely sensitive to the calibration delay, especially when the number of data streams is close to the maximum number of spatial degrees of freedom. However, under certain conditions, the system is not sensitive to the calibration delay for the scalable distributed precoding used in CF-mMIMO when the number of data streams is small.

\begin{appendices}
\setcounter{equation}{0}
\renewcommand\theequation{A.\arabic{equation}}
\section{Performance loss of local precoding with the phase drift of the calibration coefficients}
We assume that there are $L$ RRUs, each RRU  has $N_{\rm A}$ antennas, and the total number of downlink data streams is $N_{\rm A}$.
Suppose that the mean of the phases of the calibration coefficients are given and that there is no amplitude mismatch, the OTA channels are perfectly known.
We also assume that the phase drift of each antenna in the same RRU is the same (see Figure \ref{fig_coef_tf_a}).
For simplicity of analysis, we consider local ZF precoding. Then, there is no interstream interference for L-ZF. The SNR loss at the receiver can be expressed as follows:
$$\delta  = \frac{1}{{{L^2}}}{\left| {\sum\limits_{n = 1}^L {{e^{\iota {\theta _n}}}} } \right|^2},$$
where
$${\theta _n} \sim \left( { - {\theta _{\max }},{\theta _{\max }}} \right).$$
We have
\begin{align}
E\left( {{{\left| {\sum\limits_{n = 1}^L {{e^{\iota {\theta _n}}}} } \right|}^2}} \right) = \sum\limits_{n = 1}^L {\sum\limits_{m = 1}^L {E\left[ {{e^{\iota \left( {{\theta _n} - {\theta _m}} \right)}}} \right]} }
\end{align}
For ${\theta _n} \ne {\theta _m}$, we have the following
\begin{align}
E\left[ {{e^{\iota \left( {{\theta _n} - {\theta _m}} \right)}}} \right] &= \frac{1}{{4\theta _{\max }^2}}\int_{ - {\theta _{\max }}}^{{\theta _{\max }}} {{e^{\iota {\theta _n}}}{\rm{d}}{\theta _n}\int_{ - {\theta _{\max }}}^{{\theta _{\max }}} {{e^{ - \iota {\theta _m}}}{\rm{d}}{\theta _m}} }  \nonumber\\
&= \frac{{{{\sin }^2}\left( {{\theta _{\max }}} \right)}}{{\theta _{\max }^2}}.
\end{align}
Then, we obtain
$$\delta  = \frac{{{{\sin }^2}\left( {{\theta _{\max }}} \right)}}{{\theta _{\max }^2}} + \frac{1}{L}\left[ {1 - \frac{{{{\sin }^2}\left( {{\theta _{\max }}} \right)}}{{\theta _{\max }^2}}} \right].$$

In the prototype system, the phase drift range is $(-\pi/6, \pi/6)$. According to the above result, the SNR loss is approximately 0.3 dB, and the averaged SE loss is approximately 0.1 bps/Hz at high SNRs.
\end{appendices}

\bibliographystyle{IEEEtran}

\bibliography{citations}

\end{document}